\documentclass[10pt,fleqn]{article}
\usepackage{multicol,graphicx}
\begin{document}
\sf

\vspace{9mm}

\begin{center}
{\Large\sffamily\bfseries Low scale gravity as the source
of neutrino masses?}
\end{center}

\vspace{2mm}
\centerline{\large Veniamin Berezinsky,
Mohan Narayan,
Francesco Vissani
}

\centerline{\em INFN, Laboratori Nazionali del Gran Sasso,
I-67010 Assergi (AQ), Italia}
\vspace{6mm}

\centerline{\large  Abstract}
\begin{quote}
\small 
We address the question whether low-scale gravity alone
can generate the neutrino mass matrix needed to accommodate 
the observed phenomenology.
In low-scale gravity the 
neutrino mass matrix in the flavor basis 
is characterized by one 
parameter (the gravity scale $M_X$) and by an exact or approximate 
flavor blindness (namely, all 
elements of the mass matrix are of comparable size). 
Neutrino masses and mixings are consistent with the 
observational data for certain values of the matrix elements,
but only when the spectrum of mass is inverted or degenerate.
For the latter type of spectra  
the parameter $M_{ee}$ probed in double beta  
experiments and the mass parameter probed by cosmology 
are close to existing upper limits.
\end{quote}
\rm 



\section{Motivations and context}
\label{motivation}
The see-saw mechanism \cite{see-saw} remains most attractive one for 
generation of neutrino masses. The neutrino masses induced by quantum 
gravity are widely discussed, too \cite{planck1,planck2,planck3,vnb}, but 
the absolute value 
of neutrino mass $m_{\nu} \sim v^2/M_{\rm Pl}\sim 
2.5\times 10^{-6}$~eV, where $v=174$~GeV is
electroweak vacuum expectation value and   
$M_{\rm Pl}=1.2\times 10^{19}$~GeV is the Planck mass, is too small
to fit the the observational data. This  mass term is most naturally 
responsible for subdominant effects~\cite{vnb,BNV1}.

The general approach to gravity-induced neutrino mass consists in
following. 

The unknown quantum gravity Lagrangian is assumed to be expanded at
low energies in series of non-renormalizable operators, each being
inversely proportional to the powers of the Planck mass: 
\begin{equation}
{\cal L}(\psi,\phi) =
\frac{{\cal O}(1)}{M_{\rm Pl}}\ \psi\psi\phi\phi
+\frac{{\cal O}(1)}{M^2_{\rm Pl}}\ \psi\psi\psi\psi+ ...~~,
\label{expans}
\end{equation}
where $\psi$ and $\phi$ are fermion and boson fields, respectively.\\
The inverse
proportionality to $M_{\rm Pl}$ is a natural condition of vanishing of
these operators when $M_{\rm Pl} \to \infty$, i.e. when gravity is
switched off. Assuming the coefficients ${\cal O}(1)$ in expansion 
(\ref{expans}) we follow argumentation of Hawking \cite{hawking}. 
Quantum gravity Lagrangian 
${\cal L}$ and the operators of its expansion (\ref{expans})
should break the global symmetries \cite{hawking,planck1}. 
It could be understood, for
example, as absorbing a global charge by virtual black hole with its
consequent evaporation. In particular, naively one may expect that
these operators could be  flavor blind. On the other hand
these operators should respect the gauge symmetries and gauge discrete
symmetries.  In particular, the Lagrangian (\ref{expans}) must have 
SU(2)$\times$U(1) symmetry for the Standard Model fields, before this
symmetry is spontaneously broken.

In this paper we address the question whether the gravity in
extra-dimension theory with the fundamental scale $M_X < M_{\rm Pl}$
can provide an alternative mechanism for generation of neutrino masses. 

There are two specific features in gravitationally induced 
neutrino masses. The first one is gravity scale $M_X$, the second 
is flavor blindness. The scale is essentially unknown and 
{\em a priori} can be in
the range $1~{\rm TeV} \leq M_X \leq M_{\rm Pl}$. The future
development of extra-dimension theory and observations can determine  
$M_X$, and neutrino masses hopefully give now the first indication for 
this value. It is tempting to assume that flavor blindness is {\em exact}
in gravity-induced operators, and the ratios of the coefficients in 
expansion (\ref{expans}) are exactly 1. However, as we argue below the
case of approximate flavor blindness is more natural.
The exact values of these 
coefficients can be known in the framework of explicit theory of
quantum gravity. In this paper we shall analyze the both cases of exact
and approximate flavor blindness.
The above signatures of the discussed mechanism are not very promising,  
though they are not worse than for the see-saw mechanism. 

What is preferable scale of gravity in extra-dimension theory? 
A fundamental result of this theory is that the scale can be less
than the Planck mass. The real break-through in the status of 
this theory was made in works \cite{extr-dim}, where it was demonstrated
that the scale can be as low as ${\cal O}$~(TeV), without
contradiction with basic observational data. An attractive feature of 
this version is the solution of the hierarchy problem without imposing 
the supersymmetry. However, TeV gravity scale meets severe
problems with much higher scale constraints, coming from different 
elementary particle processes, such as $\mu \to e\gamma$, 
$\pi \to e\nu$, $K_1$ - $K_2$ mass difference, proton decay and 
neutrino masses. The different symmetries should be imposed 
(see e.g. \cite{zurab}) to forbid these processes. Thus it seems quite
plausible that the scale is much larger, as follows from the above-mentioned 
processes, and can reach $M_X \sim 10^{15}$~GeV, where according
Horava-Witten scenario \cite{hw} gravity starts to feel the extra dimensions. 
In our work we shall use such a large scale for the numerical analysis.
More generally, neutrino mass can provide us with the first reasonable 
indication to the fundamental gravity scale $M_X$, if it 
1~TeV$ \ll M_X <M_{\rm Pl}$.
For the gravity-induced neutrino mass, the scale responsible for it
is strictly fixed as the {\em physical quantity}, 
{\em i.e.}~it is the fundamental gravity scale $M_X$.

This approach to the theory of neutrino masses 
can be compared with the more conventional one based on 
GUT ideas. It seems that GUT models are superior
since they have their own motivations 
and happen to produce neutrino masses in the correct ball-park.
But at closer examination, this argument is not completely 
satisfactory. For instance, it is 
possible to achieve supersymmetric SU(5) unification 
with $M_{\rm GUT} \sim 2\times 10^{16}$~GeV, but firstly 
SU(5) {\em by itself} cannot be a theory of right-handed 
neutrino masses, since $\nu_R$'s are gauge singlets and secondly
the scale $M_{GUT}$ is anyway one-two order of magnitude too large 
to provide the ``observed'' neutrino masses.
On general grounds, SO(10) is more appealing, since it hosts  $\nu_R$.
However, there are (at least) two types of SO(10) models: 
those where the $\nu_R$ mass come from non-renormalizable 
coupling with $\overline{16}_H\cdot \overline{16}_H/M_{Pl}$ and
those where the $\nu_R$ mass is provided by  
renormalizable coupling with $\overline{126}_H$ higgs.
In the first case, the $\nu_R$ mass is once again induced by
the fundamental gravity scale, 
as evident from the presence of $1/M_{Pl}$ term.
In the second case, the scale of $\nu_R$ mass can be either the 
scale of left-right (intermediate) symmetry breaking, or a 
scale arising accidentally (see e.g.~\cite{ggg}).
{}From this brief examination, two conclusions can be derived: 
{\em (i)}~The {\em detailed} GUT model is needed to fix the physical 
meaning of the mass scale responsible for neutrino masses. 
{\em (ii)}~In some popular SO(10) models, one must resort 
again to quantum gravity. At present there is no selfconsistent 
``standard'' GUT model, where neutrino masses are numerically predicted  
on the basis of internal GUT scale with clear physical meaning and 
fixed value.  The advantage of the GUT theories is  the principal
possibility to perform the analytic model calculations, while 
such possibility does not exist yet in quantum gravity.

Now, in order to provide the connection with neutrino masses,
we offer a brief overview of the present 
experimental and theoretical situation.
The magnitude of the neutrino masses and mixings are governed by the
texture of the neutrino mass matrix in the flavor basis, 
denoted by $M$.
This is related to the diagonal mass matrix 
via the relation,
$M=U^*\; {diag}(M)\; U^\dagger$, where $U$ is the usual
neutrino mixing matrix specified by three angles,
$\theta_{12}=\omega$, $\theta_{23}=\psi,
\theta_{13}=\phi$ and
one CP violating phase $\delta$. The two Majorana phases $\rho$ and
$\sigma$ can be incorporated in the diagonal masses.\footnote{We assume
three light left handed Majorana neutrinos throughout the analysis.}
Neutrino oscillation data provide us with 
information on the mixing angles, but 
constrain only on the mass squared differences defined by,
$\Delta M_{21}^2 = M_2^2 - M_1^2$ and $\Delta M_{32}^2 =M_3^2 - M_2^2$ 
\cite{fstvis,crefog,smirbah} and {\it not}
on the absolute neutrino masses, though there are upper bounds coming from
laboratory experiments \cite{trit,hm} and 
from cosmology \cite{wmap}. It 
should be borne in mind that the limits coming from 
cosmology are crucially dependent on 
various assumptions \cite{disc}.
In addition, there is
no constraint at present on $\delta$ and on the Majorana phases.
This has the consequence that $M$ is not uniquely determined and there
are various textures consistent with present data.  Attempts 
to reconstruct the mass matrix using available data from neutrino
experiments are presented in \cite{alfug,fas}. However, any
phenomenological approach has to face the limitations 
outlined above.  The theoretical counterpart of this 
situation is that it is possible to 
postulate several textures of mass matrices   
which are consistent with present data. There are a large 
number of studies where this approach has been developed,
for example see \cite{jwfv,frzing,moha,fty,arss,twy,kk,
 wett,tan,gcb,perez,abjs,viss,fgm,jez}.  

A striking feature of most of the textures listed in the above references
is that there are always some 
entries which are very small or zero in the mass matrix, while 
some elements are
${\cal O}(1)$. 
This could be due to some underlying
symmetries or selection rules, see e.g., \cite{petbarb}. 
In this assumption, the texture is far from what can
be called as a ``democratic'' structure. However, 
imposing discrete symmetries on
the mass matrix can lead to a democratic structure 
for the mass matrix \cite{ppppp,brsil,magraj,kubo,grimus,has}. 
The idea of ``anarchy'' in the structure of
the mass matrix has also been investigated \cite{hmw,king}.
A nice summary of the above issues has been recently 
given in \cite{soq}.

\section{Neutrino mass textures induced by gravity}
The relevant gravitational dimension-5 operator for the spinor 
$SU(2)_L$ isodoublets,\footnote{Here and everywhere below we 
use Greek letters $\alpha,\beta,$... for the flavor states 
and Latin letters $i,j,k$... for the mass states.} 
$\psi_\alpha=(\nu_\alpha,\ell_\alpha)$ 
and the scalar one, $\varphi=(\varphi^+,\varphi^0)$, 
can be written with the operators 
introduced by Weinberg \cite{wein} as:
\begin{equation}
{\cal L_{\rm grav}} = \frac{\lambda_{\alpha \beta}}{2 M_X}
({\psi}_{Aa\alpha}\: \epsilon_{AC}\: \varphi_C)\: C_{ab}^{\scriptscriptstyle -1}\:
({\psi}_{Bb\beta}\: \epsilon_{BD}\: \varphi_D) +h.c.,
\label{grav}
\end{equation}
where $M_X$ is the gravity scale, which in the case of extra dimensional
theories can be less than $M_{\rm Pl}$ 
and $\lambda_{\alpha \beta}$ are numbers ${\cal O}(1)$. 
In eq.(\ref{grav}),
all indices are explicitly shown:
the Lorentz indices $a,b=1,2,3,4$ 
are contracted with the charge conjugation matrix $C$,
the $SU(2)_L$ isospin indices $A,B,C,D=1,2$ 
are contracted with $\epsilon=i\sigma_2$
($\sigma_m$ with $m=1,2,3$ are the Pauli matrices).
After spontaneous electroweak
symmetry breaking, the Lagrangian~(\ref{grav})
generates terms of neutrino mass:
${\cal L_{\rm mass}}= \lambda_{\alpha \beta}\: {v^2}/{2 M_X}\ \nu_{\alpha}
C^{\scriptscriptstyle -1}\nu_{\beta}$,
where $v$=174~GeV denotes the vacuum expectation value.

The matrix $\lambda_{\alpha \beta}$ gives the neutrino mass matrix in
the flavor representation. An attractive assumption of the exact flavor 
blindness of quantum gravity corresponds to  the equal values of 
$\lambda_{\alpha \beta}$, e.g. $\lambda_{\alpha \beta}=1$. 
However, the flavor blindness cannot be the exact. It is broken,
though weakly, by radiative corrections. It should be broken more strongly by 
topological fluctuations (wormhole effects).

Even in the case quantum gravity itself provides equal coupling constants in 
the Lagrangian (\ref{grav}) e.g. $\lambda_{\alpha \beta}=1$, the
topological fluctuations at the Planck scale lift this flavor
symmetry, making the coupling constants different 
\cite{coleman},\cite{giddings}. This effect can be described as
the renormalization due to the Planck-size baby universes 
which contain the appropriate particle states. In other words, 
``the ungauged coupling constants can be transferred to baby
universes'' \cite{coleman}.   
If in the parent universe all couplings are equal $\lambda_{\alpha\beta}=1$,
in the state with baby universes these 
coupling constants can be substantially different.   

It is natural to expect that the wormhole effects give to the 
Lagrangian (\ref{grav}) the contribution of the same order as  other 
mechanisms of quantum gravity, e.g. exchange by virtual black hole. 
It makes an assumption of an approximate 
flavor blindness $\lambda_{\alpha \beta} \sim {\cal O}(1)$,
used in the earlier works \cite{planck1} - \cite{planck3}, 
plausible.  In the applications below we shall study the both cases, 
exact and approximate flavor blindness, demonstrating that exact
flavor blindness as the mechanism for neutrino-mass generation is 
disfavoured.

\subsection{Exact flavor blindness}
Let us consider first the case of the exact flavor blindness
and show that it cannot describe all observational data.

The mass matrix in the flavor basis is given by:
\begin{equation}
M =  \mu \left( \begin{array}{ccc}
  1 & 1 & 1 \\
  1 & 1 & 1 \\
  1 & 1 & 1 
\end{array} \right), 
\label{textm}
\end{equation}
were we have defined
$\mu = {v^2}/{M_X}$. 
Since the eigenvalues of the matrix are $3,0,0$ in units of $\mu$, it is
obvious that there is only one scale for oscillations. Let us first
demand that this is the solar neutrino scale. Equating the scale obtained
from eq.(\ref{textm}) to the present best fit value of 
$7 \times 10^{-5}$~eV$^2$ obtained
from analyses of
neutrino data \cite{fstvis} gives us:
\begin{equation}
M_X \approx 10^{16}~\mbox{GeV} .
\end{equation}
The texture specified in eq.(\ref{textm}) gives the following form for
the neutrino mixing matrix:
\begin{equation}
U = \left( \begin{array}{ccc}
  \frac{1}{\sqrt{2}} & \frac{1}{\sqrt{6}} & \frac{1}{\sqrt{3}} \\
  0 & -\sqrt{\frac{2}{3}} & \frac{1}{\sqrt{3}} \\
  -\frac{1}{\sqrt{2}} & \frac{1}{\sqrt{6}} & \frac{1}{\sqrt{3}} 
\end{array} \right). 
\label{ufdm}
\end{equation}
With this texture we have only one non-zero mass difference 
$\Delta M^2=M_3^2-M_1^2=M_3^2-M_2^2$, which can be identified with 
$\Delta M_{\rm sol}^2$. It is straightforward to
calculate the survival probability $P_{ee}$ and $\theta_{\rm sol}$ as 
$\sin^2 2\theta_{\rm sol}=8/9$, or $\theta_{\rm sol} \approx 35^\circ$ 
in good agreement with data \cite{crefog,smirbah}.
The value $U_{e3} = {1}/{\sqrt{3}}$ is compatible with 
CHOOZ constraints \cite{chooz} 
since in this case $\Delta M^2_{sol} \langle L/E \rangle \ll 1$.

Alternatively, requiring the oscillation scale to  
be the atmospheric neutrino scale of 
$2-3 \times 10^{-3}~\mbox{eV}^2$, we get:
\begin{equation}
M_X \approx 10^{15}~\mbox{GeV} .
\end{equation}
But here the CHOOZ constraint \cite{chooz}
becomes operative and since the texture in eq.(\ref{textm})
predicts $U_{e3}=1/\sqrt{3}$, it is observationally excluded.
So there is no space to explain the atmospheric neutrino problem.

\vspace{2mm}
Hence we see that an exact  
flavor blind texture, generated at 
a typical 
GUT
scale can explain at best the solar neutrino problem
in terms of oscillations. An additional mechanism is needed 
to provide the atmospheric neutrino mass squared difference 
(even though it is not clear whether such a mechanism can be 
implemented without destabilizing the value 
of the solar mixing angle that we found).

\subsection{Approximate flavor blindness}
In the rest of this work we shall consider the case of an approximate 
flavor blindness $\lambda_{\alpha\beta}$.
We will follow a straightforward procedure
to compare this assumption with the data.
First we construct the diagonal mass 
matrix allowed by the data, using the known values 
of $\Delta M^2$'s and the bounds on the absolute mass scale.
Then we transform the mass matrix to  
the flavor basis, using the neutrino mixing matrix that
satisfies the observational constraints. 
Finally, we select the cases when all elements of the obtained
mass matrix are ${\cal O}(1)$.
 
\paragraph{General form of the mass matrix} 
Let us consider first the general form of neutrino mixing matrix. In 
the approximation $\theta_{13}=0$ it has a form
\begin{equation}
U = \left( \begin{array}{ccc}  c_{\omega} & s_{\omega} & 0 \\
-\frac {s_{\omega}}{\sqrt{2}} & \frac {c_{\omega}} {\sqrt{2}} &
 \frac {1} {\sqrt{2}} \\ 
\frac {s_{\omega}}{\sqrt{2}} & -\frac {c_{\omega}} {\sqrt{2}} &
 \frac {1} {\sqrt{2}}  
\end{array} \right), \label{defU}
\end{equation}
where $s_{\omega} = \sin \omega$ and $c_{\omega} = \cos \omega$ with
$\omega$ being the solar neutrino mixing angle.
The atmospheric neutrino mixing angle $\psi=\theta_{23}$ is taken
to be $45^\circ$.
The mass matrix in the diagonal form is taken to be:
\begin{equation}
{diag}(M) = \mbox{diag}(z_1,z_2,z_3),
\label{moe}
\end{equation}
where $z_1= M_1 e^{2 i \rho}$ and $z_2 = M_2 e^{2 i \sigma}$ are complex
numbers, $z_3=M_3$ is real and $\rho$ and $\sigma$ denote the Majorana phases.
We now transform ${diag}(M)$ to the flavor basis using 
the matrix given in eq.(\ref{defU}) and obtain:
\begin{equation}
M = \left(\begin{array}{ccc}  
z_1 c^2_{\omega}+z_2 s^2_{\omega} 
& \frac{1}{2 \sqrt{2}} s_{2 \omega} 
(z_2 -z_1 ) 
& -\frac{1}{2 \sqrt{2}} s_{2 \omega}(z_2 -z_1) 
\\[3ex]
\frac{1}{2 \sqrt{2}} s_{2 \omega}(z_2 -z_1) &
\begin{array}{l}
\frac{1}{2}(z_1 s^2_{\omega}+z_2 c^2_{\omega} 
+z_3) 
\end{array}
&
\begin{array}{l}
\frac{1}{2}(z_3-z_1 s^2_{\omega}
-z_2 c^2_{\omega}) 
\end{array}
\\[3ex]
-\frac{1}{2 \sqrt{2}} s_{2 \omega}(z_2 -z_1) &
\begin{array}{l}
\frac{1}{2}(z_3-z_1 s^2_{\omega} 
-z_2 c^2_{\omega}) 
\end{array}
&
\begin{array}{l}
\frac{1}{2}(z_1 s^2_{\omega} 
+z_2 c^2_{\omega}+z_3) 
\end{array}
\end{array} \right). \label{mfqdm}
\end{equation}
We consider the texture specified in eq.(\ref{mfqdm}) 
for the three possible types of the
neutrino mass spectrum: hierarchical, inverted hierarchical and
degenerate.

\paragraph{Hierarchical mass spectrum}
Let us begin with the case of the hierarchical mass spectrum. If
$M_1 \ll M_2 \ll M_3$, then it is obvious from eq.(\ref{mfqdm}) 
that the elements of the 2-3 block of the matrix $M$ 
are large in comparison to the other elements.
The ${\cal O}(1)$ coefficients depart from equality 
by about one order of magnitude.
Hence, this texture is rather different from a ${\cal O}(1)$
texture and it is not compatible with the properties of the 
operators induced by gravity.

\paragraph{Inverted hierarchical spectrum}
Without loss of generality we can choose
$ M_3 \ll M_1 \approx M_2$ where 
$M_2$ and $M_1$ are split by the solar scale $\Delta M^2_{21}$.
In this case we get,
\begin{equation}
\begin{array}{l}
M =\mu \left(\begin{array}{ccc}  
e^{2 i \rho} c^2_{\omega}+ e^{2 i \sigma} s^2_{\omega} & 
\frac{1}{2 \sqrt{2}} s_{2 \omega}(e^{2 i \sigma}-e^{2 i \rho}) &
-\frac{1}{2 \sqrt{2}} s_{2 \omega}(e^{2 i \sigma}-e^{2 i \rho}) \\
\frac{1}{2 \sqrt{2}} s_{2 \omega}(e^{2 i \sigma}-e^{2 i \rho}) &
\frac{1}{2}(e^{2 i \rho} s^2_{\omega}+ e^{2 i \sigma} c^2_{\omega}) & 
-\frac{1}{2}(e^{2 i \rho} s^2_{\omega}+ e^{2 i \sigma} c^2_{\omega}) \\ 
-\frac{1}{2 \sqrt{2}} s_{2 \omega}(e^{2 i \sigma}-e^{2 i \rho}) & 
-\frac{1}{2}(e^{2 i \rho} s^2_{\omega}+ e^{2 i \sigma} c^2_{\omega}) & 
\frac{1}{2}(e^{2 i \rho} s^2_{\omega}+ e^{2 i \sigma} c^2_{\omega})
\end{array} \right), \label{mfqdminvh}
\end{array}
\end{equation}
where $\mu = \sqrt{\Delta M^2_{31}}$.
For arbitrary values of the phases, $M$
does not have an ${\cal O}(1)$ texture;
e.g., this does not happen if $\rho=\sigma$ since 
in this case some elements of the mass matrix vanish.
However, for certain values
of the phases, e.g., $\rho = 0^\circ$ and $\sigma = 90^\circ$,
eq.(\ref{mfqdminvh}) becomes:
\begin{equation}
M =\mu \left(\begin{array}{ccc}  
{\scriptstyle \cos\! 2 \omega} & 
-\frac{ \sin\! 2 \omega }{\sqrt{2}} &  
\frac{ \sin\! 2 \omega}{\sqrt{2}} \\
-\frac{\sin\! 2 \omega}{\sqrt{2}} & 
-\frac{\cos\! 2 \omega}{2} &
\frac{\cos\! 2 \omega}{2} \\
\frac{\sin\! 2 \omega}{\sqrt{2}}  & 
\frac{\cos\! 2 \omega}{2}  &
-\frac{\cos\! 2 \omega}{2} 
\end{array} \right). 
\label{ftsmf}
\end{equation}
The texture defined by eq.(\ref{ftsmf}) does have 
an ${\cal O}(1)$ structure for the large value of $\omega=34^\circ$
suggested by the data. We also have to satisfy the relation 
$\mu^2 = 2-3 \times 10^{-3}~\mbox{eV}^2 $
which results in:
\begin{equation}
M_X \approx 10^{15}~\mbox{GeV}.
\end{equation}
For the inverted hierarchy 
the mass probed in neutrinoless double
beta decay is related to the 
atmospheric neutrino mass splitting,
that is, $M_{ee} \sim \sqrt{\Delta M^2_{32}} \approx 45-55$~meV.
The case of eq.(\ref{ftsmf}) is the 
one when the two Majorana phases give origin 
to destructive interference and $M_{ee}\approx 10-20$ meV, 
but larger values can be found by varying these phases
(compare with the general discussion in \cite{fstvis}).
Thus, a large value of $M_{ee}=15-50$~meV 
characterizes the scenario where 
the spectrum of mass has an 
inverted hierarchy and a non-GUT matrix is 
responsible for the observed neutrino phenomena.

\paragraph{Degenerate mass spectrum}
For the degenerate spectrum specified by
$M_1 \approx M_2 \approx M_3 \equiv \mu$ (with the common value of
the masses $\mu$ being much more than the splittings between the levels) 
an analysis similar to the previous one 
applies. Again, an ${\cal O}(1)$ texture 
appears for certain choices of the Majorana phases.
In this case, 
for a common neutrino mass of about an electronvolt we get:
\begin{equation}
M_X \approx 10^{13}~\mbox{GeV},
\end{equation}
For the same choices of Majorana phases 
of eq.(\ref{ftsmf}) the 
mass probed in neutrinoless double beta decay is:
$M_{ee} \approx \mu \cos\! 2 \omega $, 
where $\mu$ is bounded above by the
kinematic limit coming from tritium experiments.
As in the previous case, this is the {\it minimum}
value of $M_{ee}$: other choices of 
Majorana phases will always lead to a higher value.
Of course, from the viewpoint of double beta decay experiments
this is the most appealing feature of this scenario.

\subsection{Minimal deviations from flavor blindness\label{mfv}}

As we discussed above, the approximate flavor blindness is a natural
option, and in order to describe the neutrino-mass data one should 
introduce deviations from exact flavor blindness. This brings us to the
question what is the {\em minimal} deviation which is needed to fit
the data. In practice, we will consider the  
absolute values $|M_{\alpha\beta}|$ and 
discuss, in the three cases considered above, 
for which choice of the Majorana phases the differences 
between the matrix elements are minimal.

Assuming {\bf normal hierarchy} of the spectrum we get
\begin{equation}
|M_{e x}| \le |M_{x x}|/6
\ \ \mbox{ where } x=\mu\mbox{ or }\tau,
\label{prt}
\end{equation}
where the numerical coefficient $1/6$ is the ratio of masses
$(\Delta M^2_{sol}/\Delta M^2_{atm})^{1/2}$,
or the upper bound on the mixing angle $\theta_{13}$.
The minimal deviations are anyway large and this is 
the reason why normal hierarchy is disfavored 
when we assume that all the elements of the 
mass matrix have the comparable values.

Assuming {\bf inverted hierarchy} we get
\begin{equation}
\left\{
\begin{array}{l}
|M_{ee}|^2\approx  |2 M_{x x}|^2\propto 1-\sin^22\theta_{12}\ \sin^2\xi\\
|M_{e x}|^2\propto 1/2\ \sin^22\theta_{12}\ \sin^2\xi
\end{array}
\right.
\end{equation}
where $x=\mu$ or $\tau$ as in eq.~(\ref{prt}), 
$\xi=\rho-\sigma$, and 
where the proportionality 
constant is $\sqrt{\Delta M^2_{atm}}
\approx 50$~meV;
smaller terms order $\Delta M^2_{sol}/\Delta M^2_{atm}$ are 
omitted. 
Requiring that $|M_{e x}|$ is between 
$|M_{e e}|$ and $|M_{x x}|$ we thus get the condition 
\begin{equation}
|M_{ee}|=
(0.58-0.82) \sqrt{\Delta M^2_{atm}} =29-41\mbox{ meV},
\end{equation}
which implies that it will be observable by next generation 
experiments. However, the minimal deviation 
from exact blindness is a factor of 2 (namely, the ratio between 
$|M_{ee}|$ and $|M_{xx}|$) and this 
makes the scenario of inverted hierarchy 
less appealing.

\begin{figure}[t]
\begin{center}
\includegraphics[width=8cm,angle=270]{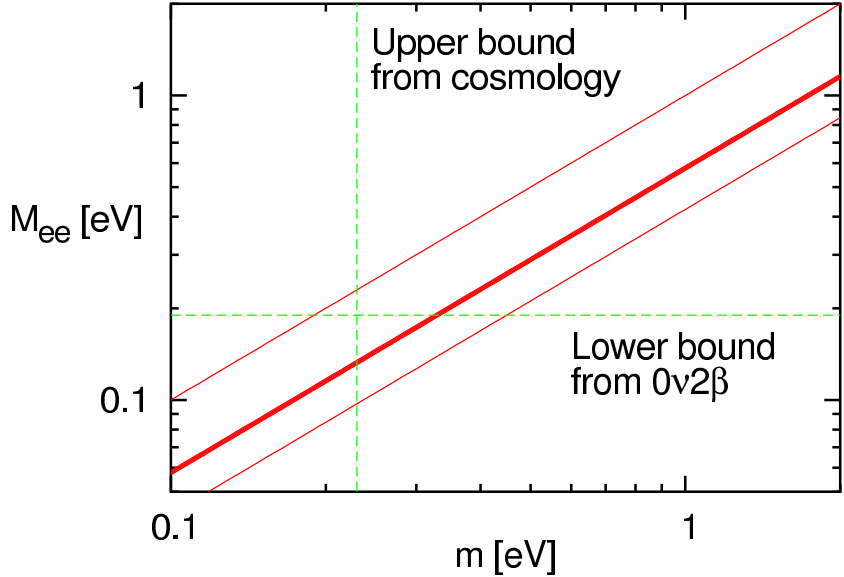}
\caption{\em Comparison of the prediction of eq.~(\ref{ssss})
(thick line) with the general range for $M_{ee}$ allowed with 
3 neutrinos (thin lines).
The upper bound on $m$ is from Ref.~\cite{wmap}. 
The lower bound on $M_{ee}$ is the 1~$\sigma$ range in Ref.~\cite{H-M}
divided by 2 in order to account for the uncertainty in 
nuclear matrix elements for the $0\nu2\beta$ transition 
${}^{76}{\rm Ge}\to {}^{76}{\rm Se}\ e^{-}\ e^{-}$.
\label{fig}}
\end{center}
\end{figure}

Assuming {\bf degenerate neutrinos}, we obtain the most interesting
scenario. In fact, when we fix the Majorana 
phases $\rho$ and $\sigma$, prescribing  $|M_{ee}|=|M_{e\mu}|$ and 
$|M_{\mu\tau}|=|M_{\mu\mu}|$, we immediately realize that
it is possible to arrange
\begin{equation}
|M_{\alpha\beta}| = m/\sqrt{3},\ \ \mbox{ where }\alpha,\beta=e,\mu,\tau
\label{ssss}
\end{equation}
setting again $\theta_{13}=0$ and $\theta_{23}=45^\circ$,
and omitting the small $\Delta M^2_{ij}/m^2$ terms.

This implies that the limit obtained 
in cosmology, $m<0.23$~eV at 95 \% CL
can be translated in a limit on 
the mass seen in double beta decay:
\begin{equation}
|M_{ee}|<0.13\ \mbox{eV}
\end{equation} 
This is smaller than the value 
suggested by the new analysis of the Heidelberg-Moscow data~\cite{H-M}
$0.44_{-0.20}^{+0.14}$~eV (at 3 $\sigma$ level),
even if one takes $1 \sigma$ lower experimental 
value and assumes that the nuclear matrix elements have been 
underestimated  (see figure \ref{fig}). 
Such a small value $m_{\nu_e}\approx m\approx 0.23$~eV, being
interesting for the forthcoming KATRIN experiment,
does not not guarantee a 3~$\sigma$ discovery 
with the present facility.
From eq.~(\ref{ssss}) one  obtains the value of the mass 
$M_{ee}=0.19$~eV   
\begin{equation}
m_{\rm cosm}= 3 m=1.0 \mbox{ eV},
\end{equation}
where we used $M_{ee}=0.19$~eV, which is 
1~$\sigma$ lower than the 
value that explains the Heidelberg-Moscow results, 
reduced by 50 \%.
This implies that, in deriving cosmological  bound  
the errors were underestimated or the assumptions do not hold. 
A value $m_{\nu_e}\approx m\approx 0.33$~eV
should permit a 3~$\sigma$ discovery in KATRIN.

Note that when we assume that the violation 
of flavor blindness is minimal, 
the theoretical framework becomes more restrictive 
and the predictions for 
the neutrinoless double beta decay process become 
more precise. This can be understood well  
from figure~\ref{fig}, since for instance one can reconcile 
the cosmological bound and Heidelberg-Moscow findings 
in the 3 neutrino context if $M_{ee}\sim m\sim 0.2$~eV.
Of course, under this assumption 
one has the non-minimal deviations from the 
flavor blindness.

The case of eq.~(\ref{ssss}) has been considered 
for the first time by Frigerio and Smirnov in a phenomenological
analysis of all possible neutrino mass matrices: 
see ref.~(\cite{fas}), second paper. Our
approach permits a step forward, since this is not one case among many
other, but the only possible case once that we require that the deviations
from flavor blindness are minimal.

\section{Renormalization group effects}
As we have seen in the previous section, the scales where we want gravity
to generate the mass matrix are typical GUT scales or a little lower.
Hence one has to consider the effect of renormalization 
group (RG) evolution from this scale to
the electroweak scale on the mass matrix.
In both the standard model and its
minimal supersymmetric extension, the RG effects are negligible as far as the
structure of the mass matrix is concerned. However oscillation observables
can be significantly altered, especially for the degenerate mass spectrum.
For the parameter space of interest at present to oscillation 
phenomenology, modulo fine tuning, the effects of RG are 
not very large\footnote{The conditions 
when the RG effects can be treated as a perturbation are 
discussed in ref.\cite{vnb}.} 
\cite{vnb,pok,lind,arc} except for very 
large values of the common neutrino mass (that as recalled 
above are not favored by the data) and of the 
parameter $\tan\! \beta$ (but only for supersymmetric models).

\section{Summary and discussion}
We have considered dimension-5 gravity induced operators, 
suppressed by a low scale of gravity 
$M_X<M_{\rm Pl}$,  as the source of the neutrino mass matrix.
After spontaneous electroweak symmetry breaking with
$\langle\varphi^0\rangle=v$, the operator in eq.(\ref{grav})
produces the neutrino mass matrix in flavor basis
$M_{\alpha\beta}=\lambda_{\alpha\beta}\; v^2/M_X$. 
We assume an approximate flavor blindness when all 
$\lambda_{\alpha\beta} \sim {\cal O}(1)$.
The exact values of $\lambda_{\alpha\beta}$ is a prerogative of the detailed
theory of quantum gravity and the wormhole theory. 
 
We have demonstrated that for the degenerate
and the inverse hierarchical neutrino mass spectrum there are sets of 
${\cal O}(1)$ coefficients when the neutrino masses and mixings satisfy
all the observational data. The mass scale for this case is
$M_X \sim 10^{13} - 10^{15}$~GeV, i.e., close to a typical GUT scale. 
The discussed mechanism is not valid for 
the hierarchical neutrino mass spectrum.

In the extreme case of unbroken flavor blindness, $\lambda_{\alpha\beta}=1$,
the gravity-induced neutrino mass matrix can explain at best the
solar-neutrino data and an additional mechanism is needed 
to provide the atmospheric neutrino mass squared difference.

The gravity-induced textures have interesting predictions
for the mass $M_{ee}$ probed in neutrinoless double 
beta decay. 
For the inverted hierarchical neutrino mass spectrum the predicted 
mass parameter $M_{ee}$ is large, 15 - 50~meV. For degenerate spectra, 
$M_{ee}$  can saturate the bound from cosmology of 0.23~eV or 
possibly approach the bound of $\sim 1$~eV coming from the studies 
of tritium endpoint spectra. 
In both cases it can explain the result
obtained from the recent analysis of the Heidelberg-Moscow 
experiment, although the range of compatibility 
between this result and cosmology is restricted.

The conclusions of previous paragraph are more generic than 
that related to gravity-induced neutrino mass model. The specific feature 
of this model is approximate flavor blindness which favors degenerate
mass spectrum of neutrinos. Another consequence of this model is the
further narrowing of the compatibility of cosmological bounds and 
the Heidelberg-Moscow results.  As discussed in Sect.~\ref{mfv}, this is 
due to the fact that the Majorana phases are {\em fixed} by the condition 
of flavor-blindness: see eq.~(\ref{ssss}) and fig.~\ref{fig}.
Approximate flavor blindness is a natural prediction of the
gravity-induced  neutrino mass model, 
but not the exclusive signature of this model. This model can be
probed only by combination of the numerical value of fundamental 
gravity scale, found from other data, and approximate flavor
blindness. At present stage of development, 
we can only argue that this model is a
viable possibility. 

\section*{Acknowledgement}
We are grateful to V. Rubakov and A. Vilenkin for very helpful
discussion of quantum gravity and wormhole effects. 

\vskip1cm
\footnotesize 
\frenchspacing
\begin{multicols}{2}

\end{multicols}

\end{document}